\documentclass[preprint,showpacs,preprintnumbers,aps,pre]{revtex4-1}
\begin{document}

\title{Retainability of canonical distributions for a Brownian particle controlled by a time-dependent harmonic potential}
\author{Geng Li}
\affiliation{Department of Physics, Beijing Normal University, Beijing 100875, China}
\author{Z. C. Tu}\email[Corresponding author. Email: ]{tuzc@bnu.edu.cn}
\affiliation{Department of Physics, Beijing Normal University, Beijing 100875, China}

\begin{abstract}
The retainability of canonical distributions for a Brownian particle controlled by a time-dependent harmonic potential is investigated in the overdamped and underdamped situations, respectively. Because of different time scales, the overdamped and underdamped Langevin equations (as well as the corresponding Fokker-Planck equations) lead to distinctive restrictions on protocols maintaining canonical distributions. Two special cases are analyzed in details: First, a Brownian particle is controlled by a time-dependent harmonic potential and embedded in medium with constant temperature; Second, a Brownian particle is controlled by a time-dependent harmonic potential and embedded in a medium
whose temperature is tuned together with the potential stiffness to keep a constant effective temperature of the Brownian particle.
We find that the canonical distributions are usually retainable for both the overdamped and underdamped situations in the former case. However, the canonical distributions are retainable merely for the overdamped situation in the latter case. We also investigate general time-dependent potentials beyond the harmonic form and find that the retainability of canonical distributions depends sensitively on the specific form of potentials.


\end{abstract}

\maketitle

\section{Introduction\label{Sec-one}}

In statistical mechanics, canonical distribution is a crucial and universal distribution function for equilibrium systems in contact with thermal reservoirs. This kind of distribution function ensures the equipartition of energy of the system, based on which we may easily define an effective temperature of the system in the sense of statistical mechanics. In equilibrium system, the defined effective temperature is exactly equal to the temperature of thermal reservoir~\cite{Reichl1998}. If the distribution function of a system maintains canonical form when it departs from equilibrium, the equipartition of energy of the system still holds. Therefore, we may define the effective temperature of this non-equilibrium system in analogy with the equilibrium situation.

On the other hand, most systems discussed in classical thermodynamics are in equilibrium or quasi-static states where we need not mention the distribution functions of systems. A famous example is the Carnot cycle, which consists of two isothermal and two adiabatic processes.
The efficiency of ideal Carnot heat engine, so called the Carnot efficiency, gives the upper bound of efficiencies of heat engines operating between two thermal reservoirs. Since the isothermal processes are in quasi-static states, we may expect that the distribution function of the heat engine (the working substance) follows canonical distribution in each isothermal process. The realization of quasi-static process requires infinite duration of the Carnot cycle while the work done by engine in each cycle is finite. Thus the power output of the Carnot heat engine is vanishing, which limits the practical applications of Carnot heat engines. To achieve finite power output, the cycle should be accelerated. This requirement gives rise to the development of finite-time thermodynamics~\cite{Chambadal1957,Novikov1957,Curzon1975}.

One of the key topics in finite-time thermodynamics is the issue of efficiency at maximum power.
The well-known Curzon-Ahlborn efficiency $\eta_{CA}\equiv 1-\sqrt{T_{c} / T_{h}}$ was phenomenologically derived for an endoreversible heat engine operating at maximum power~\cite{Curzon1975}, where $T_{h}$ and $T_{c}$ represent the temperatures of the hot reservoir and cold reservoir, respectively.
This result has attracted much attention from physicists and engineers for many years~\cite{Andresen1977,Hoffmann1985,Vos1985,Chen1989,Chen1994,Bejan1996,Broeck2005,HernandezPRL07,Tu2008,Esposito2009,Esposito2010,Gaveau2010,Wang2012EPL,Wang2012PRE,Izumida2012,Wang2012a,Wang2012,Guo2013,Sheng2014,Sheng2015,OuerdaneEPJS,RocoEPJST,Schmiedl2008,Seifert2012,Tu2014}. Recently, two stochastic models~\cite{Schmiedl2008,Tu2014} of heat engines have been proposed to discuss the availability of the Curzon-Ahlborn efficiency from the level of statistical mechanics. Schmiedl \textit{et al}~\cite{Schmiedl2008} constructed a stochastic heat engine by using a time-dependent harmonic potential to control a Brownian particle. With the consideration of the overdamped situation, they found that the efficiency at maximum power of their stochastic heat engine is slightly smaller than the Curzon-Ahlborn efficiency. In the isothermal processes of their heat engine, the distribution functions of the Brownian particle keep canonical (Gaussian form for harmonic potential) if they are initially canonical. Recently, one of the present authors~\cite{Tu2014} generalized their stochastic heat engine into the underdamped situation and recovered the Curzon-Ahlborn efficiency exactly. The canonical distributions are also retained in the isothermal processes of this model. In both models, the retainability of canonical distributions are crucial for their discussions. A natural question is: what conditions make the canonical distributions to be retainable for finite-time processes? In addition, the isothermal processes in the above stochastic heat engines are not really ``isothermal" in the traditional sense, since the effective temperature of the Brownian particle is a time-dependent variable although the temperature of the medium keeps constant in each isothermal precesses. Whether can we maintain the canonical distributions with constant effective temperatures by simultaneously tuning the potential stiffness and the medium temperature?

In this work, we aim at answering the two aforementioned questions. The rest of this paper is organized as follows. In Sec.~\ref{Sec-two}, the general equations of motion are introduced. In Sec.~\ref{Sec-three} and Sec.~\ref{Sec-four}, we study the canonical distributions (Gaussian form) of a Brownian particle controlled by a time-dependent harmonic potential in the overdamped situation and underdamped situation, respectively. We divide each sections into two cases: First, the canonical distribution corresponds to a time-dependent effective temperature while the temperature of the reservoir keeps constant; Second, the canonical distribution corresponds to a constant effective temperature while the temperature of the reservoir is tuned. In Sec.~\ref{Sec-five}, we further discuss the requirements to retain canonical distributions in general potentials.

\section{Equations of motion\label{Sec-two}}

We consider a Brownian particle moving in a one-dimensional potential. Let $x$ and $p$ represent the coordinate and the momentum of the particle, respectively. A time-dependent potential $U(x,\lambda)$ is applied on the particle, where the function $\lambda=\lambda(t)$ represents the controlling protocol. The motion of the Brownian particle may be described by the Langevin equation~\cite{Gardiner1989}:
\begin{equation}\label{eq:underdampedLE} \frac{dx}{dt}=p, \quad     \frac{dp}{dt} = - \frac{\partial U(x,\lambda)}{\partial x} - \gamma p + \xi(t), \end{equation} where $\gamma$ is the coefficient of friction and $\xi(t)$ represents the Gaussian white noise satisfying $\langle \xi(t) \rangle=0$ and $\langle \xi(t) \xi(t^\prime) \rangle=2\gamma T \delta(t-t^\prime)$. Here $T$ represents the temperature of thermal reservoir. For the sake of simplicity, we have set the mass of particle and the Boltzmann factor to be unit. If the coefficient of friction, $\gamma$, is so large that we can neglect the inertial effect, the Langevin equation will reduce to
\begin{equation}\label{eq:overdampedLE} 0 = - \frac{\partial U(x,\lambda)}{\partial x} - \gamma \frac{dx}{dt} + \xi(t). \end{equation}

The Langevin equations mentioned above are derived by eliminating the rapidly varying degrees of freedom in a more detailed microscopic model. Each Langevin equation has a suitable time resolution. The time resolution of the underdamped Langevin equation (\ref{eq:underdampedLE}) is the duration between successive collisions from the molecules of medium, which is much smaller than the damping time of inertia effect $\tau_c \equiv 1/\gamma$. However, the overdamped Langevin equation (\ref{eq:overdampedLE}) holds only for systems whose time resolutions are much larger than $\tau_c$. Different from the underdamped Langevin equation, the overdamped Langevin equation loses some information in the timescales smaller than $\tau_c$~\cite{Sekimoto1997,Sekimoto2010}. It was recently found that the overdamped Langevin equation (\ref{eq:overdampedLE}) fails to express the entropy production, while an anomalous entropy production may arise in the limit of small inertia of the underdamped case~\cite{Kim2004,Celani2012,Ge2014,Ge2015,Kawaguchi2013}. The difference that we will derive below between overdamped case and underdamped case might be related to this effect.

Corresponding to the Langevin equations, the distribution functions $\rho$ of the Brownian particle are governed by the Fokker-Planck equations~\cite{Gardiner1989},
\begin{equation}\label{eq:underdampedFPE} \frac{\partial \rho}{\partial t} = -\frac{\partial}{\partial x} (p\rho) + \frac{\partial}{\partial p}\left( \gamma p\rho + \frac{\partial U}{\partial x}\rho + \gamma T \frac{\partial \rho}{\partial p} \right) \end{equation}
in the underdamped situation and
\begin{equation}\label{eq:overdampedFPE} \frac{\partial \rho}{\partial t} = \frac{1}{\gamma} \frac{\partial}{\partial x}\left( \frac{\partial U}{\partial x}\rho + T\frac{\partial \rho}{\partial x} \right) \end{equation}
in the overdamped situation, respectively.

\section{Retainability of canonical distributions in the Overdamped situation\label{Sec-three}}
In this section, we mainly consider a Brownian particle controlled by a time-dependent harmonic potential $U(x,\lambda)=\lambda(t)x^{2}/2$ in the overdamped situation. The canonical distributions (Gaussian form) in the configuration space have been used by Schmiedl and Seifert~\cite{Schmiedl2008} when they investigated the efficiency at maximum power of stochastic heat engines. Here, we will investigate the retainability of canonical distributions in the overdamped situation.

\subsection{Time-dependent effective temperature\label{subsec-A}}

In this subsection, we consider a Brownian particle embedded in medium with constant temperature. We investigate how to maintain the canonical distribution (Gaussian form) function
\begin{equation}\label{eq:overdampedDF1} \rho(x,t) = \sqrt{\frac{\beta(t) \lambda(t)}{2\pi}} \textrm{exp}\left( - \beta(t) \frac{\lambda(t)x^{2}}{2} \right) \end{equation}
for a Brownian particle controlled by a time-dependent harmonic potential.
The above distribution function (\ref{eq:overdampedDF1}) implies that $1/\beta(t)$ could be interpreted as an effective temperature of the Brownian particle in the sense of statistical ensemble. Substituting Eq.~(\ref{eq:overdampedDF1}) into the overdamped Fokker-Planck equation (\ref{eq:overdampedFPE}), we obtain
\begin{eqnarray}\label{eq:overdampedlimit}  \frac{1}{2} \left( \dot{\beta}\beta^{-1} + \dot{\lambda} \lambda^{-1} \right) \rho - \frac{1}{2} \Big( \dot{\beta} \lambda + \beta \dot{\lambda} \Big)\rho x^{2}  =\frac{\lambda}{\gamma} \Big( 1-\beta T \Big)\rho - \frac{\beta \lambda^{2}}{\gamma} \Big( 1-\beta T \Big) \rho x^{2}.  \end{eqnarray}
In this paper, the dot on a variable represents the derivative of that variable with respect to time. Comparing the left-hand side and right-hand side of the above equation, we obtain
\begin{equation}\label{eq:overdampedRequire1} \dot{\beta} \lambda + \beta\dot{\lambda}  = 2 (1- \beta T )\frac{\beta\lambda^{2}}{\gamma}  \end{equation}
due to the arbitrariness of coordinate $x$.

Equation (\ref{eq:overdampedRequire1}) is a first order ordinary differential equation for any given protocol $\lambda=\lambda(t)$. We can usually find its solution $\beta=\beta(t)$ such that the distribution functions remain canonical all the time if they are initially canonical. Thus the canonical distribution with time-dependent effective temperature is retainable in the overdamped situation.

\subsection{Constant effective temperature\label{subsec-B}}
In the above subsection, we have discussed the retainablitiy of canonical distributions with time-dependent effective temperature for the Brownian particle.
We wonder if it is possible to realize a truly ``isothermal'' process for the Brownian particle, where the effective temperature is kept constant by tuning the potential stiffness $\lambda(t)$ and medium temperature $T(t)$, simultaneously.

When the effective temperature is constant, the canonical distribution (Gaussian form) function becomes
\begin{equation}\label{eq:overdampedDF2} \rho(x,t)= \sqrt{\frac{\beta_{0} \lambda(t)}{2\pi}} \textrm{exp}\left( - \beta_{0} \frac{\lambda(t)x^{2}}{2} \right) \end{equation}
with $1/\beta_{0}$ being the constant effective temperature. It is not hard to verify that the above distribution function is retainable if
\begin{equation}\label{eq:overdampedRequire2} \gamma \dot{\lambda} = 2[1-\beta_{0} T(t)]\lambda^{2} \end{equation}
is fulfilled. From Eq.~(\ref{eq:overdampedRequire2}), we obtain
\begin{equation}\label{eq:overdampedDF2b}
T(t)=\frac{1}{\beta_0}\left[1-\frac{\gamma\dot{\lambda} }{ 2 \lambda^2}\right].
\end{equation}
That is, for any given protocol $\lambda=\lambda(t)$, if the temperature of medium is tuned according to Eq.~(\ref{eq:overdampedDF2b}), the canonical distribution (\ref{eq:overdampedDF2}) with constant effective temperature $1/\beta_0$ is retainable in the overdamped situation.
Therefore, the truly ``isothermal'' process could be realized by tuning $T(t)$ and $\lambda(t)$ according to Eq.~(\ref{eq:overdampedDF2b}).

\section{Retainability of canonical distributions in the Underdamped situation\label{Sec-four}}
In this section, we mainly consider a Brownian particle controlled by a time-dependent harmonic potential $U(x,\lambda)=\lambda(t)x^{2}/2$ in the underdamped situation. The canonical distributions (Gaussian form) in the whole phase space have been used by one of the present authors~\cite{Tu2014} when investigating the efficiency at maximum power of underdamped stochastic heat engines. Here, we will investigate the retainability of canonical distributions in the underdamped situation.

\subsection{Time-dependent effective temperature\label{subsec-A}}
Now we consider a Brownian particle embedded in medium with constant temperature and investigate how to maintain the canonical distribution (Gaussian form) function
\begin{equation}\label{eq:underdampedDF1} \rho(x,p,t) = \frac{\sqrt{\beta^{2}(t)\lambda(t)}}{2\pi} \textrm{exp}\left[ -\beta(t)\left( \frac{p^{2}}{2}+ \frac{\lambda(t)x^{2}}{2} \right) \right] \end{equation}
for the Brownian particle controlled by a time-dependent harmonic potential.
Here, $1/\beta(t)$ could be interpreted as an effective temperature of the Brownian particle.
Substituting Eq.~(\ref{eq:underdampedDF1}) into the underdamped Fokker-Planck equation (\ref{eq:underdampedFPE}), we obtain
\begin{eqnarray}\label{eq:underdampedlimit} \left(   \dot{\beta}\beta^{-1} + \frac{1}{2} \dot{\lambda}\lambda^{-1} \right) \rho - \frac{\dot{\beta}}{2} \rho p^{2} - \frac{1}{2}\left( \dot{\beta} \lambda + \beta \dot{\lambda} \right)\rho x^{2}  = \gamma \left( 1-\beta T \right)\rho - \gamma\beta \left( 1-\beta T \right) \rho p^{2}. \end{eqnarray}
Considering the arbitrariness of momentum $p$ and coordinate $x$, we derive
\begin{equation}\label{underdampedRequire1} \beta \lambda = \textrm{const} \end{equation}
and
\begin{equation}\label{underdampedRequire2} \dot{\beta} = 2 \gamma \beta (1-\beta T).  \end{equation}

Solving Eqs.~(\ref{underdampedRequire1}) and (\ref{underdampedRequire2}), we obtain
\begin{equation}\label{protocol1}  \beta(t)= \frac{\beta_{0}}{\beta_{0} T + (1-\beta_{0} T) e^{-2 \gamma t}} \end{equation}
and
\begin{equation}\label{protocol2} \lambda(t)= \lambda_{0} \left[ \beta_{0} T+(1-\beta_{0} T)e^{-2\gamma t} \right] \end{equation}
where $\lambda_{0}\equiv\lambda(0)$ and $\beta_{0}\equiv \beta(0)$ are the initial values of $\lambda$ and $\beta$ at $t=0$, respectively.
Thus, in the underdamped situation, if we try to maintain the canonical distribution in a time-dependent harmonic potential, there is no extra freedom to choose the protocol except for (\ref{protocol2}). This feature is quite different from that in the overdamped situation.

\subsection{Constant effective temperature\label{subsec-B}}
Here we consider another case that the effective temperature is kept constant by tuning the potential stiffness $\lambda(t)$ and medium temperature $T(t)$, simultaneously.

Substituting the canonical distribution function
\begin{equation} \rho(x,p,t) = \frac{\sqrt{\beta_{0}^{2}\lambda(t)}}{2\pi} \textrm{exp}\left[ -\beta_{0}\left( \frac{p^{2}}{2}+ \frac{\lambda(t)x^{2}}{2} \right) \right] \end{equation}
with constant effective temperature $1/\beta_{0}$ into Fokker-Planck equation (\ref{eq:underdampedFPE}), we derive
\begin{equation} \lambda(t) = \textrm{const} \end{equation}
and
\begin{equation} T(t)={1}/{\beta_{0}},  \end{equation}
which rules out any nontrivial protocol that can retain canonical distributions.

Therefore, in the underdamped case, it is impossible to retain the canonical distributions with constant effective temperature by simultaneously tuning the potential stiffness $\lambda(t)$ and the medium temperature $T(t)$. This feature is also quite different from that in the overdamped situation.

\section{General potentials beyond the harmonic form\label{Sec-five}}

In the above discussions, we have investigated the retainability of canonical distributions for a Brownian particle in a time-dependent harmonic potential. Here we extend our discussion to general time-dependent potentials beyond the harmonic form.

\subsection{The overdamped situation\label{subsec-A}}
First, we consider a Brownian particle embedded in medium with constant temperature $T$ in the overdamped situation. The canonical distribution function in a general potential $U=U(x,\lambda(t))$ has the form
\begin{equation}\label{eq:overdampedGDF} \rho(x,t) = Z_{o}^{-1} e^{- \beta(t) U} \end{equation}
with $Z_{o} \equiv \int e^{- \beta(t) U} dx$ being the partition function. Substituting it into the overdamped Fokker-Planck equation (\ref{eq:overdampedFPE}), we obtain
\begin{eqnarray}\label{eq:overdampedGP}  \left\langle \dot{\beta} U + \beta \dot{\lambda} \frac{\partial U}{\partial \lambda}  \right\rangle - \left(  \dot{\beta} U + \beta \dot{\lambda} \frac{\partial U}{\partial \lambda}   \right)   = \frac{1}{\gamma} \left( 1- \beta T \right) \left[\frac{\partial^{2} U}{\partial x^{2}}  - \beta \left( \frac{\partial U}{\partial x} \right)^{2} \right], \end{eqnarray}
where $\langle \cdot \rangle$ indicates the average over the canonical distribution function. Obviously, the protocols maintaining general canonical distributions depend on the specific form of the potentials.

If the potential $U$ is an analytical function of $x$, we could expand it into Taylor series at the point $x=0$ as $U(x, \lambda) = \sum_{n=0}^{\infty} a_{n}(\lambda) x^{n}$. Substituting it into Eq.~(\ref{eq:overdampedGP}), we have
\begin{eqnarray}\label{eq:overdampedGPS1} && \left\langle  \frac{\partial}{\partial t}(\beta U)  \right\rangle  \delta_{n,0} - \left(  \dot{\beta} a_{n} + \beta \dot{a}_{n} \right)  = \frac{1}{\gamma}(1-\beta T)  [ (n+2)(n+1)a_{n+2}  \nonumber \\  &-&\beta \sum_{m=0}^{n} (n-m+1)(m+1) a_{n-m+1} a_{m+1} ],~~ (n=0,1,\cdots). \end{eqnarray}
Therefore, for a general potential, we should tune the potential stiffness $\lambda(t)$ according to Eq.~(\ref{eq:overdampedGPS1}) to maintain the canonical distributions for a Brownian particle embedded in medium with a constant temperature. If the potential is of harmonic form, $a_{2}= \lambda/2$ and $a_{n} = 0$ for $n \ne 2$, the requirement (\ref{eq:overdampedGPS1}) will return to Eq.~(\ref{eq:overdampedRequire1}) derived above. For the other potential forms, Eq.~(\ref{eq:overdampedGPS1}) will give a relation between $\lambda$ and $\beta$ for each $n$. Different relations might be contradictory with each other, which rules out nontrivial protocols that retain canonical distributions.

Next, we consider the case to retain the canonical distributions with a constant effective temperature by tuning the medium temperature $T(t)$.
The canonical distribution function could be written as
\begin{equation}\label{eq:overdampedGDF2} \rho(x,t) = Z_{o}^{-1} e^{- \beta_{0} U} \end{equation}
with $Z_{o} \equiv \int e^{- \beta_{0} U} dx$ being the partition function. Substituting it into the overdamped Fokker-Planck equation (\ref{eq:overdampedFPE}), we obtain
\begin{eqnarray}\label{eq:overdampedGP2} \beta_{0} \dot{\lambda} \left\langle   \frac{\partial U}{\partial \lambda}  \right\rangle -    \beta_{0} \dot{\lambda} \frac{\partial U}{\partial \lambda}     = \frac{1}{\gamma} \left[ 1- \beta_{0} T(t) \right]\left[ \frac{\partial^{2} U}{\partial x^{2}}  - \beta_{0} \left( \frac{\partial U}{\partial x} \right)^{2}\right].  \end{eqnarray}
Expanding $U(x, \lambda)$ as Taylor series, $U(x, \lambda) = \sum_{n=0}^{\infty} a_{n}(\lambda) x^{n} $, and substituting it into Eq.~(\ref{eq:overdampedGP2}), we obtain the following requirement
\begin{eqnarray}\label{eq:overdampedGPS2} && \beta_{0} \left\langle  \frac{\partial U}{\partial t}  \right\rangle  \delta_{n,0} -  \beta_{0} \dot{a}_{n}  = \frac{1}{\gamma}[1-\beta_{0} T(t)]  [ (n+2)(n+1)a_{n+2}  \nonumber \\  & - &\beta_{0} \sum_{m=0}^{n} (n-m+1)(m+1) a_{n-m+1} a_{m+1} ],~~ (n=0,1,\cdots). \end{eqnarray}
That means, for a general potential, if we want to maintain the canonical distributions with the constant effective temperature, the medium temperature $T(t)$ should be tuned according to Eq.~(\ref{eq:overdampedGPS2}).
In the case of harmonic potentials, Eq.~(\ref{eq:overdampedGPS2}) will degenerate to Eq.~(\ref{eq:overdampedDF2b}).
For the other potential forms, Eq.~(\ref{eq:overdampedGPS2}) will give a relation between $\lambda(t)$ and $T(t)$ for each $n$. Different relations might be contradictory with each other, which rules out nontrivial protocols that retain canonical distributions.

Therefore, in the overdamped situation, the existence of nontrivial protocols to retain canonical distributions mainly depends on the requirements (\ref{eq:overdampedGPS1}) or (\ref{eq:overdampedGPS2}), which strongly relies on the specific form of potentials.

\subsection{The underdamped situation\label{subsec-B}}
First, we consider a Brownian particle embedded in medium with constant temperature $T$.
In the underdamped situation, the canonical distribution function becomes
\begin{equation}\label{eq:underdampedGDF} \rho(x,t) = Z_{u}^{-1} e^{- \beta(t) ( \frac{p^{2}}{2} + U)} \end{equation}
with the partition function $Z_{u} \equiv \sqrt{{2 \pi}/{\beta(t)}} \int  e^{- \beta U} dx$. Substituting it into the underdamped Fokker-Planck equation (\ref{eq:underdampedFPE}), we derive
\begin{eqnarray}\label{eq:underdampedGP}  \frac{1}{2} \dot{\beta} \beta^{-1}  + \left\langle \dot{\beta} U + \beta \dot{\lambda} \frac{\partial U}{\partial \lambda}  \right\rangle   - \left(  \dot{\beta} U + \beta \dot{\lambda} \frac{\partial U}{\partial \lambda}   \right)  - \frac{\dot{\beta}}{2} p^{2}   = \gamma \left( 1- \beta T \right) - \gamma \beta \left( 1- \beta T \right)p^{2}. \end{eqnarray}
Due to the arbitrariness of momentum $p$, we have
\begin{equation}\label{underdampedGRequire1}  \dot{\beta} U + \beta \dot{\lambda} \frac{\partial U}{\partial \lambda} = \left\langle \dot{\beta} U + \beta \dot{\lambda} \frac{\partial U}{\partial \lambda}  \right\rangle \end{equation}
and
\begin{equation}\label{underdampedGRequire2} \dot{\beta} = 2 \gamma \beta (1-\beta T).  \end{equation}
The above requirement (\ref{underdampedGRequire2}) determines the time-dependent behavior of $\beta=\beta(t)$. Therefore, the key point lies in the existence of solutions to (\ref{underdampedGRequire1}) for general potential $U=U(x,\lambda(t))$.

If $U$ is an analytical function of $x$, we can expand it as $U( x,\lambda) = \sum_{n=0}^{\infty} a_{n}(\lambda) x^{n}$. Then, the requirement (\ref{underdampedGRequire1}) reduces to
\begin{equation}\label{underdampedGRequire1b} \beta a_{n} = C_{n}  \end{equation}
with $C_{n}$ being a time-independent coefficient for each $n$. From Eqs.~(\ref{underdampedGRequire2}) and (\ref{underdampedGRequire1b}) we can solve
\begin{equation}  a_{n} = C_{n} \left[   T + ({1}/{\beta_{0}}- T) e^{-2 \gamma t}  \right]  \end{equation}
where $\beta_{0}=\beta(0)$. If we suppose
\begin{equation}\label{potentialstrength} \lambda(t) = \left[   1 + \left(\frac{1}{\beta_{0}T}- 1 \right) e^{-2 \gamma t}  \right]  \end{equation}
and
\begin{equation}\label{potentialform}  u(x) \equiv T\sum_{n=0}^{\infty} C_{n} x^{n}, \end{equation}
the potential may be expressed as
\begin{equation}\label{potential} U(x,\lambda)  = \lambda(t) u(x). \end{equation}
In this case, if we interpret $\lambda(t)$ as the strength of potential $U$, it is clear that we can solely alter the potential strength $\lambda(t)$ according to protocol (\ref{potentialstrength}) to maintain canonical distributions in the underdamped situation. The specific function form of (\ref{potentialform}) is irrelevant.

Now, we consider the case to maintain the canonical distributions while keeping effective temperature constant by tuning the potential stiffness $\lambda(t)$ and medium temperature $T(t)$, simultaneously. When the effective temperature is constant, the canonical distribution function becomes
\begin{equation}\label{eq:underdampedGDF2} \rho(x,t) = Z_{u}^{-1} e^{- \beta_{0} ( \frac{p^{2}}{2} + U)} \end{equation}
with the partition function $Z_{u} \equiv \sqrt{{2 \pi}/{\beta_{0}}} \int e^{- \beta_{0} U} dx$. Substituting it into the underdamped Fokker-Planck equation (\ref{eq:underdampedFPE}), we can obtain
\begin{eqnarray}\label{eq:underdampedGP2} \dot{\lambda} \left\langle  \frac{\partial U}{\partial \lambda}  \right\rangle   - \dot{\lambda} \frac{\partial U}{\partial \lambda} = \gamma \left[ \frac{1}{\beta_{0}}-  T(t) \right] - \gamma \left[ 1- \beta_{0} T(t) \right]p^{2}. \end{eqnarray}
Due to the arbitrariness of momentum $p$, we derive a trivial requirement that
\begin{equation} T(t)={1}/{\beta_{0}}.  \end{equation}
The rest of Eq.~(\ref{eq:underdampedGP2}) contains two terms with different coordinate $x$ dependence. Considering the arbitrariness of coordinate $x$, we may derive
\begin{equation} \lambda(t) = \textrm{const}. \end{equation}
Therefore, it is still impossible to realize a truly ``isothermal'' process for the Brownian particle in general time-dependent potential by tuning potential stiffness $\lambda(t)$ and medium temperature $T(t)$, simultaneously.

\section{Conclusion and discussion\label{Sec-six}}

In this paper, we discussed the retainability of canonical distributions for a Brownian particle controlled by a time-dependent harmonic potential in the overdamped and underdamped situations, respectively. First, we considered the case to maintain canonical distributions with time-dependent effective temperature. In the overdamped situation, we could retain canonical distributions (Gaussian form) if they are initially canonical. However, in the underdamped situation, the protocols $\lambda(t)$ to control the harmonic potential should abide by Eq.~(\ref{protocol2}) if we want to maintain canonical distributions. Further, we investigated the condition to keep the constant effective temperature by varying the medium temperature and potential stiffness with time. We found that, in this condition, canonical distributions are retainable merely in the overdamped situation.

The general time-dependent potentials beyond the harmonic form were also investigated, where the retainability of canonical distributions for a Brownian particle depends sensitively on the form of potentials. In the overdamped situation, the restrictions for protocols to maintain canonical distributions in general potentials were different from the case of harmonic potentials. If the potential deviates from the harmonic form, the conclusions derived in~\cite{Schmiedl2008} need to be reconsidered. However, we could derive similar protocols to retain canonical distributions in the underdamped situation. Therefore, the conclusions derived in~\cite{Tu2014} is still available for general potentials.

The present discussions about the retainability of canonical distributions are based on the normal Langevin equation. Recently, Liu \emph{et al.} put forward a new Langevin equation by introducing stochastic noises in the whole phase space~\cite{Liu2015}. Whether it is possible to construct canonical distribution corresponding to the new form of Langevin equations will be investigated in our future work. The novel framework put forward by Ao and his coworkers~\cite{Ao2004,Ao2007,Yuan2012,Xing2010} to handle stochastic processes may provide new methods to this question.

\section*{Acknowledgement}

The authors are grateful to the helpful discussions with Ao Ping and Shiqi Sheng. This work is supported by the
National Natural Science Foundation of China (Grant No. 11322543) and the Fundamental Research Funds for the Central Universities (NO. 2015KJJCB01).

\end{document}